# General Automatic Solution Generation for Social Problems

Tong Niu[1†]    Haoyu Huang[1†]    Yu Du[1†]    Weihao Zhang[1]    Luping Shi[1,2,3]    Rong Zhao[1,2,3*]

[1] Center for Brain-Inspired Computing Research (CBICR), Optical Memory National Engineering Research Center, and Department of
Precision Instrument, Tsinghua University, Beijing 100084, China.

[2] IDG/McGovern Institute for Brain Research, Tsinghua University, Beijing 100084, China.

[3] THU-CET HIK Joint Research Center for Brain-Inspired Computing

† These authors contributed equally: Tong Niu, Haoyu Huang, Yu Du†

*e-mail: r_zhao@tsinghua.edu.cn

**Abstract:** Given the escalating intricacy and multifaceted nature of contemporary social systems, manually generating solutions to address pertinent social issues has become a formidable task. In response to this challenge, the rapid development of artificial intelligence has spurred the exploration of computational methodologies aimed at automatically generating solutions. However, current methods for auto-generation of solutions mainly concentrate on local social regulations that pertain to specific scenarios. Here, we report an automatic social operating system (ASOS) designed for general social solution generation, which is built upon agent-based models, enabling both global and local analyses and regulations of social problems across spatial and temporal dimensions. ASOS adopts a hypergraph with extensible social semantics for a comprehensive and structured representation of social dynamics. It also incorporates a generalized protocol for standardized hypergraph operations and a symbolic hybrid framework that delivers interpretable solutions, yielding a balance between regulatory efficacy and function viability. To demonstrate the effectiveness of ASOS, we apply it to the domain of averting extreme events within international oil futures markets. By generating a new trading role supplemented with new mechanisms, ASOS can adeptly discern precarious market conditions and make front-running interventions for non-profit purposes. This study demonstrates that ASOS provides an efficient and systematic approach for generating solutions for enhancing our society.

## 1 Introduction

Providing solutions to social problems or dilemmas is a pivotal objective of social science [1-5]. Depending on the causes and impact of problems, experts propose various solutions ranging from the issuance of singular regulations, such as the frequent adjustments of tax policy, to the creation of comprehensive social forms, such as the socialist society form proposed by Marx. [6-10]. With the rapid advancements in computing sciences and artificial intelligence, the field of computational social science (CSS) [11, 12] has made progress in automatically generating tailored solutions for some scenario-specific social problems, such as welfare allocation through reinforcement learning [10, 13-15], and public opinion control through cluster analysis [8]. These auto-generated solutions predominantly refine localized social intricacies within the existing social structure. However, the manifestation of a society and its underlying dynamics emerge through a confluence of both global structural and local details [16-18]. The global structure includes the existence of those essentials, such as individuals and attributes that characterize them, and the relationships between these elements. The local details pertain to the specific manifestation and content of these essentials and relationships. Thus, an effective automatic solution generation approach requires the ability to comprehensively analyze and regulate societal intricacies from a panoramic perspective that encompasses global and local aspects. Furthermore, different solution mandates distinct interfaces between algorithms and heterogeneous social descriptions. Presently, prevailing automatic generation approaches tend to be specialized solutions, meticulously tailored to precise requisites, thereby lacking the flexibility and efficiency required for broader applications. Hence, a pressing need exists for a general interface capable of standardizing diverse heterogeneity.

The agent-based modeling (ABM) stands as a fundamental research paradigm in CSS [19-21]. It provides a bottom-up perspective for analyzing our society through the computerized simulation of interactions among decision-makers (agents) and environment objects. In this paper, we report on a general automatic social operating system (ASOS) based on ABM, which is capable of generating social solutions from local to global scales across spatial and temporal dimensions (Fig. 1a) by addressing diverse solving targets of different social problem scenarios. To achieve this, ASOS firstly imitates existing social phenomena through an agent-based model represented by a unified hypergraph [22, 23]. The hypergraph formalizes the social semantics and embeds it in its programming syntax [24-26], allowing algorithms to explicitly recognize the model's global structure and local details. The core of these social semantics encapsulates meta-information that latently governs the equilibrium states and their evolutionary trajectories within the intricate fabric of the complex social system, dominating the "computation" of the society. A symbolic hybrid framework in ASOS further iteratively generates credible solutions based on the hypergraph. An important element in this framework is the integration of neural networks and computational symbols, which harmonizes the trinity of generation efficiency, adaptability, and interpretability – a delicate equilibrium that embodies the essence of ASOS. To bridge the gap between various algorithms within this hybrid framework and the heterogeneous hypergraph operations, a generalized operating protocol is ingeniously devised to standardize operations into unified vectors. Iteratively, ASOS simulates the model for solution validation. Overall,

Then, we select the international oil futures market as a tangible and representative social scenario to empirically validate the effectiveness of ASOS. The factors affecting oil futures trading are intricate, and the trading prices fluctuate frequently. To address this, we employ ASOS to automatically generate strategies aimed at mitigating anomalous and extreme price fluctuations, thereby fostering market stability. Leveraging historical data, ASOS first replicated the price trends of WTI from 2006 to 2021 through the utilization of multi-agent trading simulations. Subsequently, through multiple iterations of generation and simulation, ASOS creates an innovative non-profit stabilizing trading role, which is capable of detecting trading behaviors that may lead to price fluctuations exceeding a certain threshold and taking preemptive interventions to hedge against potential abnormal fluctuations.

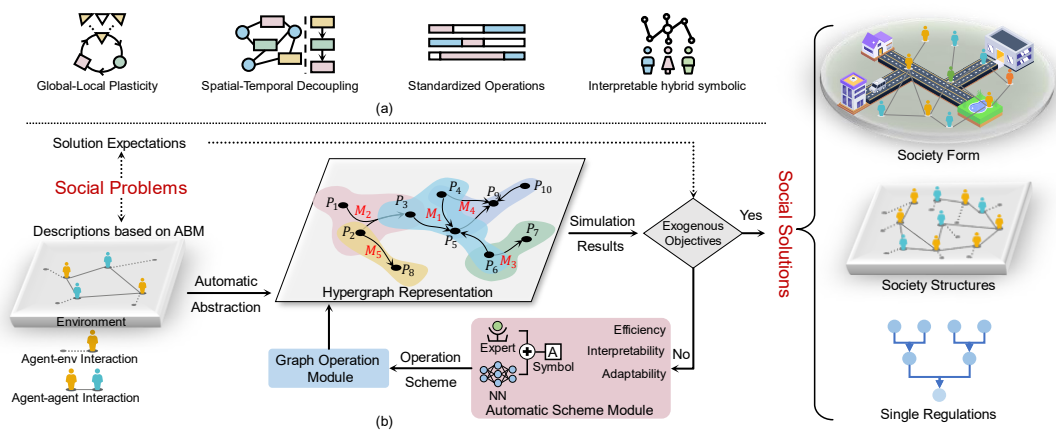

Fig. 1    Key designs and workflow of ASOS. (a) Key designs of ASOS: The global-local plasticity and spatial-temporal decoupling designs of hypergraph representations ensure the completeness and flexibility of solution generation space. The standardized operations protocol design ensures the generality of different generation algorithms. The interpretable hybrid symbolic design ensures the interpretability, efficiency and adaptability of generated solutions. (b) The workflow of ASOS.

## 2    Related work

Solution-oriented social science primarily focuses on direct resolution of practical social problems. In comparison to other purposes within social science, such as challenging conventional assumptions about the fundamental nature of social reality[27-30], providing detailed and nuanced descriptions of lived experiences[31-33], inspiring innovative modes of thinking concerning human behavior[34, 35], solution-oriented research places greater emphasis on replicability and practicality. Such research adeptly integrates insights from diverse fields, such as computational or physical science. Noteworthy studies have been conducted across different social fields, like education[36], health care[37], poverty[34] and government[38].

With the explosive growth of data and computing power over the past decade[39-41], computational social science advances unprecedently, which promotes the development of methods for social solution generation that encompass automation and intelligence. Existing related research can be broadly classified into two categories: algorithms for analyzing and addressing social problems, and modeling tools for reproducing social phenomena. In the aspect of algorithms, Zheng et al.[10] propose two level deep multiagent reinforcement learning framework that automatically designs taxation policy. Koster et al.[14] design a democratic AI that discovered mechanism that redressed initial wealth imbalance problem. Apart from applying reinforcement learning in artificial environments, many work utilize massive real-world social data to generate effective solutions, Gao et al.[42] propose a mortality risk prediction model for COVID-19 that utilizes clinical data collected at patient admission to stratify individuals based on their mortality risk. Gong et al.[43] report a novel structural-hole-based approach to control public opinion in social networks. These approaches primarily focus on addressing social problems confined to certain scenarios. In contrast, ASOS embraces a general generation approach, unconfined by the constraints of particular social scenarios.

In the aspect of modeling tools, the quality and credibility of the generated solutions are significantly influenced by modeling capacities and simulation fidelity. Therefore, various modeling platforms have been developed to provide design-friendly tools to facilitate building explicit models, such as intuitive modeling languages[25, 26, 44], user-controlled graphical panels[26, 45, 46], and even some automated model generation from scientific knowledge[47], or high-level descriptions[48]. Our work particularly emphasis on developing an extensible representation of social semantics, and achieving a balance between interpretability and programmability for both human understanding and generation algorithms.

# 3 ASOS

## 3.1 ASOS workflow

ASOS comprises three main modules: a hypergraph representation module, an automatic scheme module integrated hybrid symbolic framework, and an operation module defining the standardized operating protocol. The specific workflow (Fig. 1) is as follows: In response to a given social problem, experts first provide an initial ABM that describes existing social phenomena, along with desired outcomes for ASOS to achieve. Then, the hypergraph representation module of ASOS abstracts the initial ABM description into a unified hypergraph representation. In the subsequent step, ASOS performs the hypergraph and produces simulation results. These simulation outcomes are then combined with exogenous objectives proposed by experts to activate the scheme module (within our symbolic hybrid framework), deducing operation schemes that comply with the predefined standardized operating protocol. Moving forward, the operation module decodes schemes and applies corresponding operations to the model representation. These steps are executed iteratively until simulation results meet the stipulate objectives. Ultimately, the most optimal operation can be identified as the solution, which may range from a local refinement to an entire social form according to the operation's type and contents. The specific design of ASOS is described below.

## 3.2 Hypergraph representation with dual-plasticity for social description

As the basis for automatic social analysis and regulation, society representation with structured and extensible social semantics is the primary consideration of ASOS. Complete semantics contain static self-contained depictions of social individuals and their dynamic rules, which are dispersed in multi-class objects as agents and the environment. Therefore, considering an explicit representation that incorporates the above information, we abstract two types of general and essential representation units with unified formats: 'properties' $(P)$ that define static-state depictions of objects (Definition 1) and 'mechanisms' $(M)$ that define the dynamic process of states through the computing function that relates several specific properties (Definition 2). Therefore, mechanisms and their associated properties construct a directed hypergraph $H(P, M)$ as directed hyperedges and nodes, respectively.

**Definition 1** (Property). A property $p$ is a variable that describes static states of objects. The set of properties, $P$, is finite and non-repetitive, and defines all possible states of different classes of objects in a social model $E$. The specification of each property is shown below, where $i$ is property index, $c$ denotes the class of objects that the property depicts and $v_{\text{type}}$ is the value type used,

$$p_i \stackrel{\text{def}}{=} (i, c, v_{\text{type}}) \tag{1}$$

**Definition 2** (Mechanism). A mechanism $m$ is a function $f$ that computes the value of target properties, $P_m^{\text{target}}$, given the value of source properties, $P_m^{\text{source}}$, where $P_m^{\text{source}}, P_m^{\text{target}} \subseteq P$. Therefore, mechanism set $M$ represents the dynamics of object states over time, which is abstracted from all self-changing rules and interactive social rules of $E$. The specification of each mechanism is shown below, where $i$ is mechanism index.

$$m_i \stackrel{\text{def}}{=} (i, P_m^{\text{source}}, P_m^{\text{target}}, f),$$
$$\text{where } P_m^{\text{target}} = f(P_m^{\text{source}}) \tag{2}$$

Specifically, a property $p$ of one class that belongs to multiple instantiated objects or is used by multiple mechanisms refers to only one node. A mechanism shared among multiple objects is also declared only once, which helps to reduce the redundancy of representations significantly. In a hypergraph, the global social structure is represented by the graph topology, and local details are hidden in each node and hyperedge (Fig. 2a).

The granularity of the abstraction influences the complexity and precision of subsequent regulation. That is, the more fine-grained the hypergraph, the more precise its regulation but the greater the regulation complexity. We propose two abstraction principles to balance such trade-off, and avoid repeated logical judgments that represent social conditions to reduce computation redundancy. (Fig. 2(a)).

**Abstraction principle 1** (Independent principle). This principle forbids mechanisms from having overlapping social semantics or cross-calling. If there are dependent situations where initial expert-defined rules $r_A$ calls $r_B$ in $E$, $r_A$ will be divided into three sub-rules, the pre-call rule, $r_B$ and the post-call rule, for further abstraction.

**Abstraction principle 2** (Single responsibility principle). This principle of single responsibility divides coarse-grained mechanisms into several parts involving the minimum number of properties but ensures that each part maintains complete social semantics. This principle recommends dividing origin rules into several complete computational parts eliminating redundant logical judgments or dependencies. A complete computation involves using both the basic input and output variables solely from properties excluding intermediate variables. Dependency means that source and target properties from different mechanisms are entirely overlapped. Moreover, mechanisms should use the minimum number of properties required, in addition to satisfying the above requirement.

The specific representation and abstraction algorithms to transfer the expert' original ABM of problem scenario to the hypergraph representation is shown in supplementary information Algorithm1 and Algorithm2.

Social phenomena are influenced by social structures in the spatial domain and the sequence of events in the temporal domain. To provide plasticity from the spatial perspective, each representation unit is assigned a binary state, denoting an active or inactive state. Only active units are involved in the simulation, thereby supporting various spatial topologies for different social dynamics (Fig. 2(b)). In the temporal domain, because dynamic processes that change object states present within hyperedges, the execution sequence of the simulation is defined by the specified order of hyperedges. To provide explicit chronological dependencies of mechanisms for auto-method, this order is represented by the directed linkage between hyperedges. Hence, it is decoupled from the topology among hyperedges and nodes. We design a scheduler based on a Directed Acyclic Graph (DAG) with re-configurability to maintain these linkages, where each node represents a mechanism, and the topology order determines the execution order, thereby providing temporal plasticity (Fig. 2(c)). It also ensures a more powerful social expressiveness through supporting regulation during runtime to enable real-time intervention according to unpredictable social phenomena. The simulation process of the hypergraph representation is shown in supplementary information Fig. 1. Specifically, in consideration of the hypergraph formalism, wherein hyperedges encapsulate analogous social functions, and nodes represent singular object states, a hyperedge or a node corresponds to an object class with multiple instantiations will be executed multiple times during simulation. To obviate computing redundancy while ensuring optimal resource employment, an object allocator is devised that can propose efficient grouping strategies for categorizing instantiated objects synergistically, contingent on properties of the mechanism (supplementary information Fig. 1(a)). The cooperation between the scheduler and the object allocator achieves efficient social model simulation, effectively harnessed through the hypergraph representation (supplementary information Fig. 1(b)).

## 3.3 Hypergraph operation protocol

Using the above representation, the regulation of societies is therefore transformed into multi-scaled operations of hypergraphs [49], which are also complete because of the reconfigurability of all hypergraph parts. To support the unified deployment of these heterogeneous operations across various algorithms, the combination and standardization of such operations become imperative (Fig. 2(d)). For combination, four basic types of graph operation are adopted, tailored to the graph's scale and spatiotemporal dimensions: alteration (Fig. 2(e)), elimination, addition (Fig. 2(f)), and rescheduling (Fig. 2(g)), which refer to switching the current status of existing nodes or hyperedges, introducing exogenous properties or mechanisms, enhancing or replacing graph components, and rearranging the execution order of active mechanisms, respectively. By combining these basic operations, complex solution schemes from local to global in spatial and temporal dimensions can be implemented on hypergraphs. Then, a unified protocol is conceptualized to standardize the aforementioned operations, which stores schemes in the form of vectors. These vectors abstract the complexity of hypergraphs and can be easily generated through different intelligent algorithms. Specifically, the vectorized protocol consists of three parts: the operation type and component type, which respectively indicates whether nodes or hyperedges are subject to regulation, and the corresponding contents of the operation itself.

## 3.4 Automatic scheming with a symbolic hybrid framework

ASOS is responsible for automatically generating solution schemes that comply with the aforementioned operation protocol, aimed at adjusting artificial societies. Automatic scheming is usually realized by algorithms such as combination optimization [50] and reinforcement learning methods [10, 13, 40, 51]. However, to identify more efficient and qualified solutions for social problems or hypotheses of social phenomena, three key issues should be of concern when designing scheming algorithms. First, due to the complex nature of society, the desired objectives proposed by experts often involve various constraints and limitations, rendering them essentially multi-objective optimization problems. Therefore, there is a need to enhance the robustness and efficiency of scheme generation to navigate the potentially vast and unsmooth search space. Second, the social simulation process is dynamic. Therefore, algorithms should possess adaptability to co-evolve with changing environments. Third, the resulting scheme needs to be interpretable to provide experts with further analysis of it and the resulted social phenomenon.

To jointly address these issues, a symbolic hybrid framework is developed (Fig. 3). It contains two parts: a symbolic part and hybrid part. The symbolic part is responsible for generating interpretable symbolic expressions. It follows the framework of Cartesian genetic programming, [52-54], where an evolutionary population consists of diverse individuals that represent the variable expressions. The evolving strategy follows a manner of $\mu + \lambda$ strategy. The selection of $\mu$ and $\lambda$ can be found in supplementary information Table. 1. The specific expression comprised by symbolic operators from a pre-defined operator library (Table 1), represents a scheme that can be merged into the hypergraph as new hyperedges and nodes. Three categories of operators are developed in the operator library, namely, basic arithmetic operators, logical operators, and conditional branching operators. The flexible operator combinations support not only sequential execution, but also loop execution, which represents complex expressive ability.

The hybrid part combines expert-defined or symbols-based expressions and neural networks (NNs) [55-57] for efficient and stable enhancement of schemes. Specifically, the initially hand-crafted mechanisms serve as an excellent initial point or guidance for NN optimization, which supports efficient and reasonable emergence of enhancement schemes. NNs provide adaptability, whose output can be interpreted as a corrective term for the expert-defined expression. To fulfill it, we devise a hybrid algorithm to combine the expert-defined mechanisms (original mechanisms or expressions derived from above symbolic part) $\pi_{ed}$ and neural network mechanisms $\pi$. The input $s$ (value of source properties) are passed to both the two policies. These two types of mechanisms will both output the normal results (new value of target properties) which is network policy will learn an extra action $g(s)$ acting as a 'switch gate', thus the final output $u(s)$ can be written as

$$u(s) = g(s) \cdot \pi(s) + (1 - g(s)) \cdot \pi_{ed}(s) \quad (3)$$

where for discrete outputs, the switch gate $g(s)$ is a discrete value chosen from $\{0, 1\}$, while for continuous output the switch gate is also continuous ranging in $[0,1]$.

To train the hybrid part, a recent on-policy reinforcement learning algorithm HAPPO[58] is adopted. (The reward function is converted from the evolution objectives in the symbolic part). The training procedure is summarized in Algorithm 1. It is based on the framework of Actor-Critic[59] and CTDE (centralized training and decentralized execution)[60, 61]. For each critic, the input includes the global information and all the local information with redundant information pruned, while for each actor, the input includes only local information. The critic is only present at the training stage and dumped at execution stage. The specific algorithm is shown in Algorithm 1. The parameter details are shown in supplementary information Table 1.

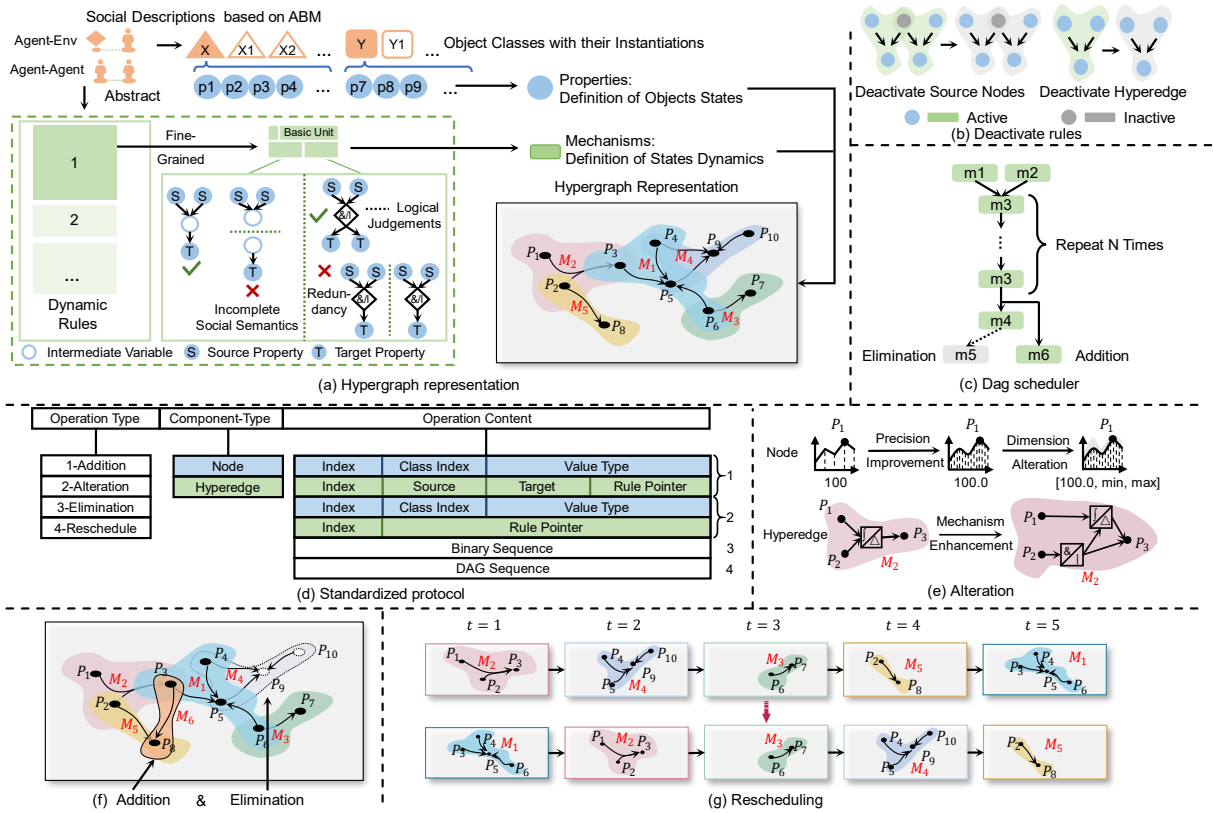

Fig. 2 Illustration of the hypergraph representation and standardized operations protocol. (a) The hypergraph representation contains social properties and mechanisms for an ABM. Orange triangles and rectangles represent different object classes (agents and environment), and hollow shapes are their instantiations. The green dashed box shows abstraction principles. (b) Switching rules of hypergraph component activation state that support spatial plasticity. Additionally, because of the computational dependencies, the deactivation of a node also deactivates the hyperedges that have it as a source node. (c) DAG scheduler that supports temporal plasticity enables flexible elimination, addition, and repetition for rescheduling. The green part of the protocol represents the operations on the hyperedge. The blue represents the operations on the node. (d) The heterogeneous operations are standardized by the protocol. There are four basic graph operations, which are alteration, addition, elimination, and rescheduling. The component types are node or hyperedge. As for operation contents, (e) alteration content: the altered class index or value type for the indicated node, or the pointer of enhanced functions for the indicated hyperedge; (f) addition content: the content of a node or hyperedge definition, elimination content: the binary activation vector for nodes or mechanisms; (g) rescheduling content: a vector that defines the execution sequence by the list of mechanism indices

# 4 Instantiation and experiments

International crude oil futures trading holds significant implications for the global economy. However, the market's multifaceted nature can lead to occasional market crashes[62, 63]. This study employs ASOS to create a model of the international crude oil futures market (ICOFM), simulate the market dynamics and generate solutions to mitigate instability.

Table 1  The symbol operator library in symbolic hybrid framework

| Symbolic Operators | Arity | Description |
|---|---|---|
| | | **Arithmetic Operators** |
| Add(x, y) | 2 | Scalar addition and element-wise vector addition. |
| Sub(x, y) | 2 | Scalar subtraction and element-wise vector subtraction. |
| Mul(x, y) | 2 | Scalar multiplication and element-wise vector multiplication. |
| Div(x, y) | 2 | Scalar division and element-wise vector division. |
| MAX($\vec{x}$) | 1 vector | Return the maximum element of $\vec{x}$. |
| MIN($\vec{x}$) | 1 vector | Return the minimum element of $\vec{x}$. |
| Max1(x) | 1 vector | $\max(x, 0)$ |
| Min1(x) | 1 vector | $\min(x, 0)$ |
| Sign(x) | 1 | $\begin{cases} 1, & x > 0 \\ 0, & x = 0 \\ -1, & x < 0 \end{cases}$ |
| x01(x) | 1 | Return $0.1x$. |
| Inv(x) | 1 | Return $1/x$. |
| Neg(x) | 1 | Return $-x$. |
| Abs(x) | 1 | Return $|x|$. |
| Sum($\vec{x}$) | 1 vector | Return the summation of all element in vector $\vec{x}$. |
| Sum3 ($\vec{x}$, $\vec{y}$, z) | 3 (vector, vector, scalar) | Return y[x[i]>=z].sum(). The sum of y's specific elements, whose corresponding elements of x are greater/less than z |
| LimUp (x,y) | 2 | $\begin{cases} 1, & x \geq y(1 + upper\_limit\_ratio) \\ 0, & x < y(1 + upper\_limit\_ratio) \end{cases}$ |
| LimDown (x,y) | 2 | $\begin{cases} 1, & x \leq y(1 - lower\_limit\_ratio) \\ 0, & x > y(1 - lower\_limit\_ratio) \end{cases}$ |
| Const0 | 0 | Constant 0 |
| Const01 | 0 | Constant 0.1. |
| Const1 | 0 | Constant 1.0. |
| Comb | n | Combine them into one list |
| | | **Logical Operators** |
| And, Or, Eq, Gt, Lt | 2 | Return the bool value through logical operation |
| Not | 1 | Return the bool value through logical operation |
| | | **Conditional Operators** |
| If-else | 1 | Return the next execution operation |
| While$_{start}$ | 1 | Starting point of while loop, next execution operation in loop |
| While$_{end}$ | 1 | Ending point of while loop, next execution operation out of loop |

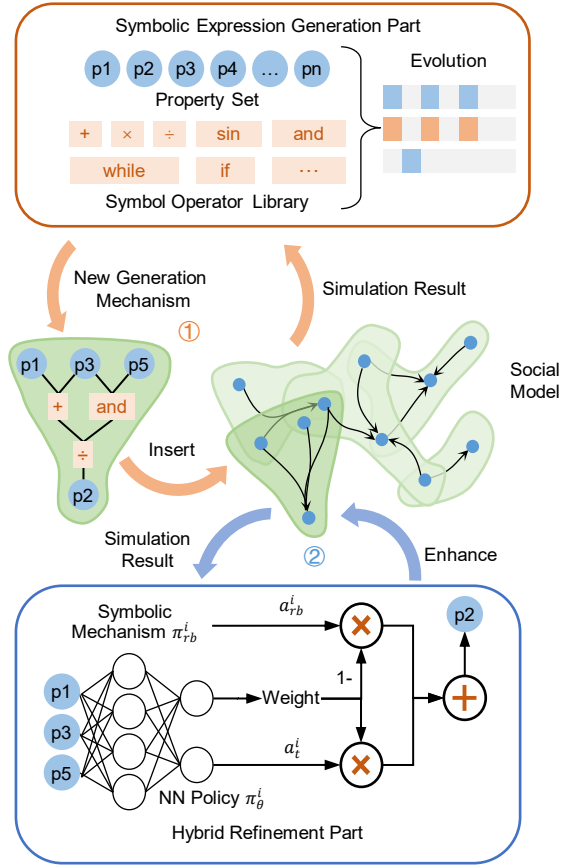

Fig. 3  The schematic of the symbolic hybrid framework. Loop ① can generate new symbolic mechanisms with the symbolic part, while the loop ② can optimize the symbolic expressions with the hybrid part. The loop ① and ② can be utilized continuously or separately.

---

**Algorithm 1** Training algorithm of hybrid part

**Require:** Initialize Rollout Buffer $\{D_i\}_{i=1}^{N}$, expert-defined rule-based policy $\{\pi_{ed}^i\}_{i=1}^{N}$, neural network policy $\{\pi_{\theta_i}\}_{i=1}^{N}$, critic $\{V_{\phi_i}\}_{i=1}^{N}$. Initialize N actor RNN states $\{h_{0,\pi}^i\}_{i=1}^{N}$ Initialize N critic RNN states $\{h_{0,V}^i\}_{i=1}^{N}$ . Max episode length T.
Reset the simulator, observe local observations $\{o_1^i\}_{i=1}^{N}$, and global states $\{s_1^i\}_{i=1}^{N}$

1) **while** step $\leq$ max_steps **do**
2)    **for** t = 1 ... T **do**
3)       **for** i = 1 ... N **do**
             $a_t^i, g_t^i, h_{t,\pi}^i = \pi_{\theta_i}(o_t^i, h_{t-1,\pi}^i)$
             $a_{ed,t}^i = \pi_{ed}^i(o_t^i)$
             $V_t^i, h_{t,V}^i = V_{\phi_i}(s_t^i, h_{t-1,V}^i)$
             $u_t^i = g_t^i \cdot a_t^i + (1 - g_t^i) \cdot a_{ed,t}^i$
4)       **end for**
5)       Execute the actions $\{u_t^i\}_{i=1}^{N}$ in the emulator, observe $\{r_t^i\}_{i=1}^{N}, \{s_{t+1}^i\}_{i=1}^{N}, \{o_{t+1}^i\}_{i=1}^{N}$.
6)       **for** i = 1 ... N **do**
7)          Store transition $(o_t^i, s_t^i, a_t^i, r_t^i, o_{t+1}^i, s_{t+1}^i, h_{t-1,\pi}^i, h_{t-1,V}^i)$ in $D_i$.
8)       **end for**
9)    **end for**
10)    **for** i = 1 ... N **do**
11)       Compute advantage estimate $\{\hat{A}_t^i\}_{t=1}^{T}$ via GAE and reward-to-go $\{\hat{R}_t^i\}_{t=1}^{T}$ on $D_i$.

| 12) | Set $M^{1:1}(s, a) = \hat{A}(s, a)$ |
|---|---|
| 13) | **end for** |
| 14) | **for** i = 1 … N-1 **do** |
| 15) | Optimize $\pi_{\theta_i}, V_{\phi_i}$ on $D_i$ w.r.t. $\theta_i, \phi_i$, with $K$ epochs, minibatch size $B$, using the advantage estimator $M^{1:i}(s, \mathbf{a})$ |
| 16) | $M^{1:i+1}(s, \mathbf{a}) = \frac{\pi_{\theta_i new}(a^i|o^i)}{\pi_{\theta_i^{old}}(a^i|o^i)} M^{1:i}(s, \mathbf{a})$ |
| 17) | **end for** |
| 18) | **end while** |

**Enterprises.** ICOFM defines three types of enterprises based on their different attributes and targets: the producer, consumer and speculator. Apart from getting more profits, the second target of the producer is to sell out the oil production within the prescribed time due to the limited storage space, while the second target of the consumer is to purchase enough oil product for satisfying the oil demand. In each simulation step, enterprises need to judge whether to submit the order and specify the order items. The enterprise perceives not only its own conditions, but also the international factors, and the time-to-time information of the futures market.

Apart from agents, the market is designed to conform to primary real market regulations: "Order", "Trading" and "Clearance and settlement".

**Order.** ICOFM provides two order types: limit order (LMT) and market order (MKT). Each order is required to specify four items:

$$\begin{Bmatrix} (contract\ month, price, quantity, side) \\ |side \in \{ask, bid\} \end{Bmatrix} \quad (4)$$

while the MKT does not need to indicate the price. Besides, ICOFM maintains an orderbook with two charts. The first chart represents the outstanding orders, which are sorted by order prices. For simplicity, quantities of contracts in buy orders are recorded as negative values, and quantities of contracts in sell orders are positive. The second chart represents the latest transactions.

**Trading.** According to the commodity futures trading rules, ICOFM contains two trading stages. Before opening the market on each trading day, the market implements the call auction. It is forbidden to submit MKT in the call auction stage. At this stage, the oil prices are set based on maximum trading volume, which will also become the opening price. The outstanding orders will be held until the market opens. After opening, the market implements the continuous double auction. At this stage, a transaction occurs if there is an outstanding order on the other side of the orderbook at a price equal to or better than the price of the submitted order. If not, the submitted order will be added to the orderbook. When multiple outstanding orders are eligible for trading, the trading priority is given in order of price first, followed by the time.

**Clearance and settlement.** ICOFM complies with the real market-to-market rule. That is, the market calculates the profit and loss of all contracts for involvers at the closing price after a day's trading, and the orderbook will be cleared on the next day. On the settle day of each futures cycle, all open contracts need to be delivered.

To ensure the credibility of our model, we incorporate real data from 2006 to 2021 [63-67] to simulate the specific international environment that influences agents' decision.

**International factors.** Crude oil futures prices in last two decades have been mainly influenced by the expectations for future oil supply and demand. Therefore, we collect the major international events that affected the oil supply and demand from 2006 to 2021 and organize them into a quarterly data set as the source of those 60 properties' values in the simulation process. According to the nature of events, these properties are divided into 5 categories (supplementary information Table 3)

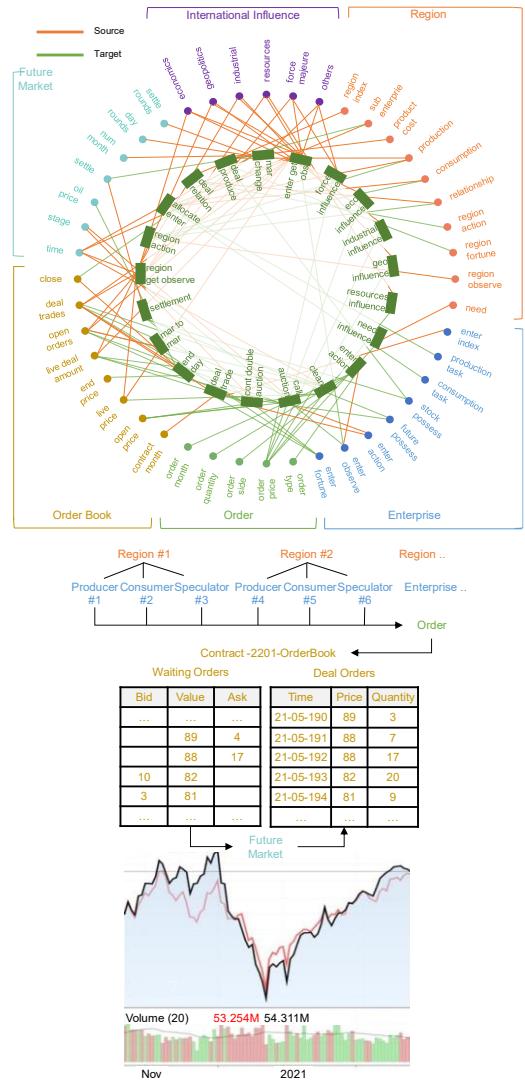

Fig. 4 The schematic of ICOFM. The above part is the ICOFM's hypergraph representation. Circles in different colors represent nodes (properties) of different object classes. Green blocks represent hyperedges (mechanisms). Orange lines represent that nodes are the source properties of hyperedges, while green lines represent that nodes are the target properties of hyperedges. The below part elaborates brifely on the relationships between different objects.

Using ASOS, the ICOFM is converted into a hypergraph. It includes 98 nodes (properties) of four object classes, which represent international factors, the trading market, regions, and enterprises, and 76 hyperedges (mechanisms), which illustrate market rules, agent trading policies, and interaction rules. The hypergraph topology, as shown in Fig. 4, implies dependencies among mechanisms and properties (where properties as international factors are categorized into five types for simplicity). The description of each hypergraph component and the specific spatio-temporal dynamic structure are shown in supplementary information Table 2-5.

## 4.2 Social phenomena reproduction

Due to the complexity of society, gaps between reality and initial social description are inevitable. To address this, the ASOS first automatically regulates the global structure and local details of the ICOFM hypergraph to enhance the credibility of future solutions, which can be referred to as social phenomena reproduction. From the global perspective, ASOS evaluates the contributions of different international factors to the fluctuation of the oil futures price. This is achieved by adjusting activation states of related properties and mechanisms using an evolutionary algorithm, ultimately determining the activation combinations of properties with minimal fitting errors over 15 years (supplementary information Table 6). The fitness is calculated as below,

$$fit_{com}^{j} = w_1 \cdot \frac{1}{N} \sum_{i=1}^{N} \left| p_{real}^i - p_{simu}^i \right| +$$

$$w_2 \cdot \sum_{i=1}^{N-1} \left| \frac{p_{real}^{i+1} - p_{real}^i}{p_{real}^i} - \frac{p_{simu}^{i+1} - p_{simu}^i}{p_{simu}^i} \right| \quad (5)$$

where $w_1$, $w_2$ are coefficients, $N$ is the number of simulation quarters, $p_{real}^i$ and $p_{simu}^i$ are the real price and simulation price in $i^{th}$ quarter. The first item indicates the absolute error between two price sequences, while the second item focuses on the increasing rate.

After refining the global structure, ASOS enhances agent behavior mechanisms through the hybrid part with reinforcement learning algorithms. For region agents, ASOS improves their production and consumption adjustment mechanisms and cooperating mechanisms, resulting in the emergence of two interesting phenomena, namely cartel[68] and co-bargaining (supplementary information Fig. 3). The emergence of the cartel phenomenon is reminiscent of the history that in 2014-2016, OPEC and Non-OPEC oil producers fell into an alliance to increase production to compete for market share under the influence of the American shale oil revolution. For enterprise agents, the ASOS improves their trading mechanisms for more profits. The specific reinforcement learning designs are as followed.

**Observation space.** Observations of enterprises and regions are same as their perceptions described in the design of regions and enterprises of ICOFM. Then, the digits of these observations are scaled to the same range $[-1, 1]$ and normalized with running mean and variance.

**Action space.** Actions of the enterprise contain whether to purpose an order, and the specific order items. The actions of the region contain the adjustment of total production, consumption and the relationship with other regions. The action space is hybrid with both discrete and continuous actions. In particular, we adopt the Beta policy[69], instead of the widely used Gaussian policy[58] for dealing with continuous actions.

**Reward design.** For enterprises, the reward functions are divided into three parts, which are the completion of the required buying and selling tasks, the completion of settlement, and their profits. The reward functions for regions contain four parts, which are the market share, the economic development, the fluctuations of oil prices, and product cost and revenue. The specific reward functions are shown in supplementary information.

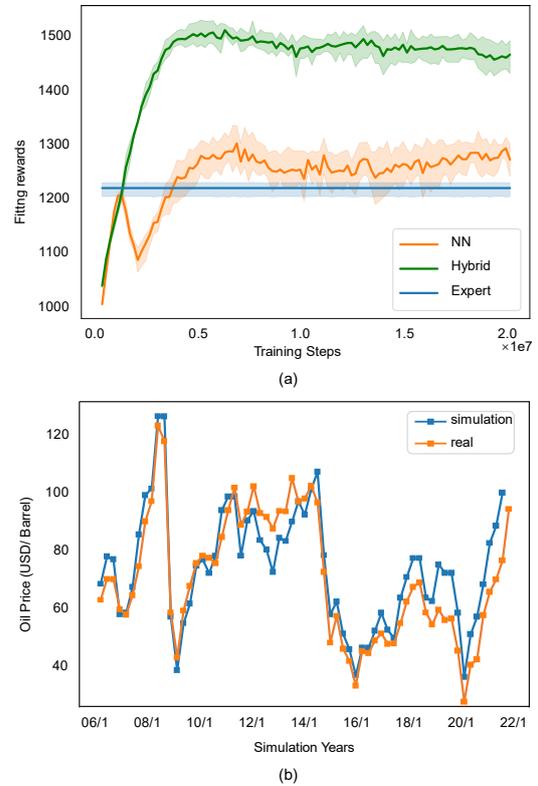

Fig.5 (a) Refining process curves of heterogeneous agents' mechanisms with hybrid part (Hybrid), pure neural networks (NN) and expert-defined policies (Expert) respectively. (b) Oil price simulation results after the refinement of global structure and local mechanisms. The blue line is the simulated oil price and the orange line is the real oil price.

The hyperparameters employed in this refinement experiment is displayed in supplementary information Table 1. The specific learning curves are shown in Fig. 5(a) and supplementary information Fig. 2, which clearly show that hybrid mechanisms outperform expert-defined mechanisms in terms of achieving higher final profits, all while exhibiting greater efficiency compared to pure NN mechanisms. The final oil futures trading simulation results are presented in Fig. 5(b).

### 4.3 Solution generation for stabilizing the futures market

Futures are originally designed to maintain value and hedge risk. However, due to information asymmetry and irrational speculative behavior, the futures market can become extremely volatile and may experience crashes. To stabilize the market, ASOS generates two single market regulations and one brand new market role through the symbolic hybrid framework.

Firstly, ASOS derives an expression that can determine if the price is above or below the fluctuation threshold through the process of evolution, which has been encapsulated and added to the fundamental operator library as LimUp/ LimDown. Then, by controlling different key market properties, ASOS generates two optimal mechanisms for preventing market collapse through the CGP evolutionary process in the symbolic hybrid framework. One mechanism involves closing the market when the fluctuation boundaries are reached, while the other restricts the trading order type (only allows "market orders" involve trading). Fig. 6 and supplementary information Fig. 4, 5 show the evolution progress for generating solutions, which includes several intermediate generated solutions and their performance rewards on both decreasing fluctuations and encouraging the dealing amount.

There are three criteria to evaluate the proposed solutions. The first criterion determines the solution's effectiveness of achieving the objective. Specifically, we use Bollinger bands to indicate the fluctuation of oil price as follows. The Bollinger bands consist of an upper band, a lower band with a moving average curve between them. Considering a simulated oil price sequence $s_t, t \in [0, T]$, where $T$ is simulation length, the N-period of moving average (MA) at time $i$

$$MA_i = \frac{1}{N} \sum_{j=i}^{i+N} p_{j,\text{simu}}, j \in [0, T-N] \quad (6)$$

is calculated, followed by the N-period standard deviation

$$\sigma_i = \sqrt{\frac{1}{N} \sum_{j=i}^{i+N} \left(p_{j,\text{simu}} - MA_i\right)^2} \quad (7)$$

where $N$ is the window length. The upper band is above the MA at $K$ times, $B_{\text{up}} = MA + K\sigma$ and the lower band is below the MA at $K$ times, $B_{\text{low}} = MA - K\sigma$. The wider bands at time $t$ indicates the bigger fluctuations at period of $[t-N, t]$. Therefore, the bands' width from $t=0$ to $t = T-N$ are averaged and taken as the overall fluctuation indicator during whole simulation periods,

$$B_{\text{avg}} = \frac{1}{T-N} \sum_{i=1}^{T-N} B_{i,\text{up}} - B_{i,\text{low}} \quad (8)$$

Besides, the increasing of trading quantity during simulation is encouraged. Therefore, the average trading quantity per quarter is taken into the final criteria as

$$fit = \lambda_1 \cdot B_{\text{avg}} + \lambda_2 \cdot \frac{1}{T} \sum_{t=1}^{T} n_{t,\text{fu}} \quad (9)$$

where the $n_{t,\text{fu}}$ is quantity of futures in each simulation step. The objective is to maximize this fitness.

If multiple proposed-solutions satisfy the first criterion, an assessment of their structural complexity $c_s$ (the second criterion) is conducted. This complexity measure is determined by multiplying the number of nodes $n_{node}$, and the average out-degree $OD_{avg}$ of the symbolic operation-based solution. Therefore, the second criterion states that solutions with lower levels of structural complexity are considered more favorable.

$$c_s = n_{node} * OD_{avg} \quad (10)$$

At last, in instances where the performance scores for the first two criteria are identical, expert intervention is required to manually determine whether a solution exhibits real-world interpretability, constituting the third criterion. Based on the assessments of the first two criteria, solution ⑤ in Fig. 6 outperforms the others. Additionally, it fulfills the third criterion as well. Therefore, we assert that it represents the most optimal social mechanism evolved to date.

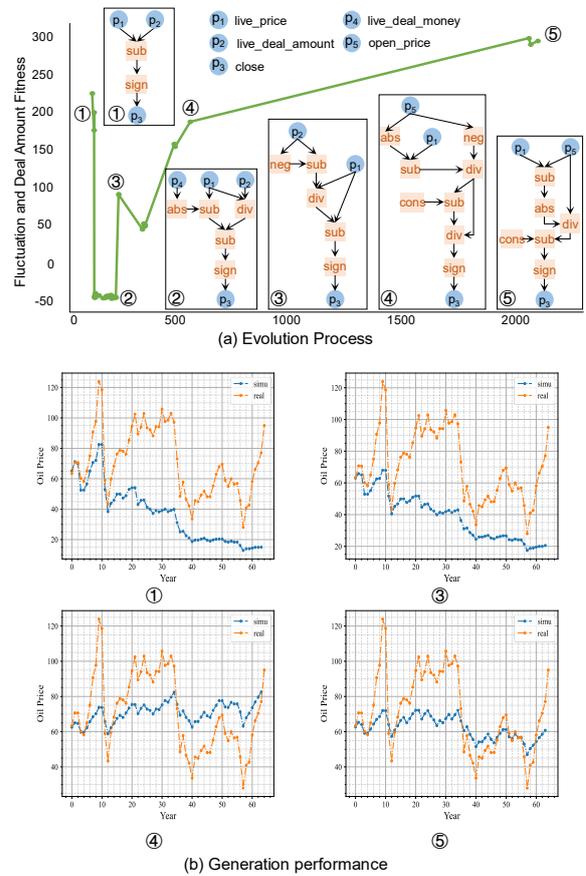

Fig. 6 The generation of "close" mechanism solution. (a) The evolving process of CGP to search the expected mechanisms in prescription experiments and several intermediate revolved results (①-⑤) during the evolving process. The 'fluctuation and deal amount fitness' is measured by the average of Bollinger bands and the deal amount. (b) The price (USD/barrel) fluctuation performance (blue line) of several solution in (a). Orange lines in each subfigure are the real price fluctuation in history.

Secondly, the ASOS generates a new trading role through adjustments of global structures and local mechanisms. This new role can be regarded as a non-profit front-running interventionist that continuously monitors the market status and real-time incoming orders. In the event that the execution of new orders (i.e., new ask and bid orders shown in Fig. 7(a)) might cause the live-time price beyond the fluctuation range, the intervener is poised to mitigate the risk of excessive price fluctuation by providing priority orders in advance. These priority orders are meticulously devised to address existing pending orders that are likely to be processed by the incoming new order at a safe price. During periods of market stability, the intervener will liquidate its positions at the original contract price. The generated "trading role" bears similarities to various established strategies in real financial markets, such as "Trend-following strategies" and "Mean reversion strategies". In the generating process, the symbolic hybrid framework first derives the logic structure to determine the necessity of submitting an intervention order, accomplished through symbols. Subsequent to this, the adjustment of prices and quantities of priority intervention orders is undertaken, performed by the NN during the generating process. Specifically, in the evolution (and hybrid part refinement) process, properties associated with "order" and "market" are adopted as source properties (observation) of the new solution. The target properties (action space) exclusively derive from the "order" object as the role attains its target through providing priority orders. The evolution is jointly driven by multiple objectives (reward function). First, the interventionist is mandated to preserve neutrality, that is to minimize the number and duration of orders held, whether bid or ask orders. Second, the interventionist's orders must not cause the real-time price to surpass the reasonable range.

In the experiment, this range is 90%-110% of the opening price. Most importantly, the number of times the real-time price breaches the fluctuation threshold during the entire simulation period is subject to minimization. The interventionist strategy schematic is presented in Fig. 7(a)-(b), and its performance on stabilizing the oil market is shown in Fig. 7(c).

## 5 Conclusions

In summary, we have developed ASOS, a tool for generating social solutions automatically to social problems, which involves a synergistic global-local cooperation in both spatial and temporal dimensions. Our study illustrated that ASOS can effectively address a wide range of diverse and intricate social problems with complex scenarios, as showcased through the demonstration of ICOFM. The general and extensible features of our generator make it applicable across various fields that extend beyond the scope of social science, encompassing realms such as ecology, industry, and medicine. A notable feature of our approach is the implementation of a standardized operation protocol, which decouples the algorithms and model representation. This design fosters the development and validation of more efficient and high-quality generative algorithms, thus bestowing a heightened capability upon this field. Our study makes a valuable contributes to the advancement of automatic research and facilitates a paradigm shift in the exploration of society phenomena.

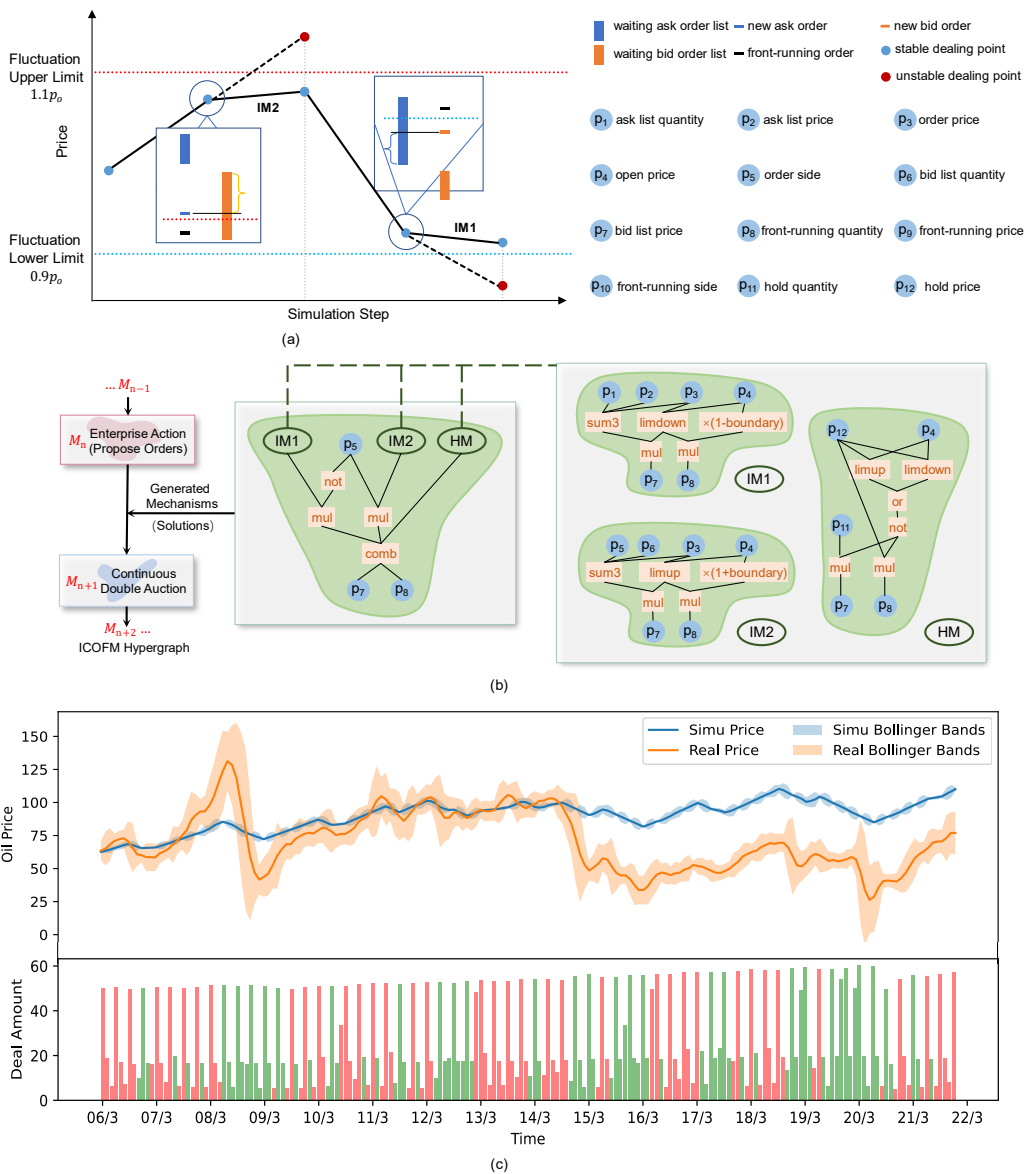

(a)

(b)

(c)

Fig 7   (a) The schematic of the trading performance of the interventionist's front-running strategy. The dashed line and red dot in the figure indicate that the absence of intervention will cause the real-time price to exceed the predetermined fluctuation range, i.e., 90% to 110% of the opening price $p_0$ (which means the boundary is 10%). The orange and blue braces in (a) denote the front-running order proposed by the intervention agent. The generated IM1 expression will be activated when an ordinary order is below the lower price bound. The generated IM2 expression will be activated when the order is to exceed the upper bound. (b) The symbolic expressions of the interventionist. There is one main function, combining three sub expressions, IM1, IM2, and HM. The main function is responsible to collect front-running orders provided by the interventionist under three conditions. IM1 and IM2 are used to determine if the incoming new order will cause the live-time price to exceed the fluctuation range. If so, IM1 and IM2 will generate priority orders, whose quantity is the sum of waiting orders that are able to be delt by the incoming new orders (realized by operator Sum3), and price is the fluctuation threshold. HM is responsible to provide orders to close positions when these orders will not lead to exceeding the fluctuation range. (c) The performance of the "non-profit front-running interventionist" solution. It ensures market dynamics and prevents extremes while conforming to the actual upward and downward trends of the market. The solid lines represent the oil price, and the shadow represents the fluctuation range of the oil price. The red and green bars represent the deal amount. The red represents the oil price increases while the green represents the oil price decreases.

# Acknowledgements

Funding: This work was partly supported by the National Key Research and Development Program of China (grant no. 2021ZD0200300), National Nature Science Foundation of China (nos. 61836004 and 62088102), IDG/McGovern Institute for Brain Research at Tsinghua University.

# Declarations of Conflict of interest

The authors have no competing interests to declare that are relevant to the content of this article.

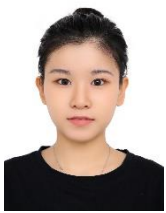

**Tong Niu** received the B.E. degree in computer science and technology from Shandong University, Shenyang, China, in 2020. She is currently a Ph.D degree candidate in instrumentation science and technology with the Department of Precision Instrument, Tsinghua University, Beijing, China. Her current research interests include computational social science, intelligent multi agent-based modeling, and brain-inspired computing.

E-mail: nt20@mails.tsinghua.edu.cn

ORCID iD: 0009-0002-6550-8764

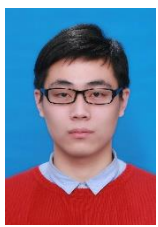

**Haoyu Huang** received the B.Sc. degree in electronic and information engineering from Xidian University, China in 2019. Currently, he is a Ph.D degree candidate in instrument science and technology at Department of Precision Instrument, Tsinghua University, China. His research interests include brain-inspired computing, machine learning and computational social science.

E-mail: huanghy19@mails.tsinghua.edu.cn

ORCID iD: 0009-0003-9794-3450

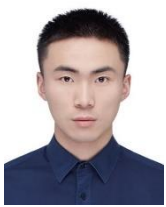

**Yu Du** is currently pursuing the Phd's degree with the Department of Precision Instrument, Tsinghua University, Beijing, China. He received the B.S. degree in Mechanical Engineering from Tsinghua University, Beijing, China, in 2020. His current research interests include foundation model, multi-modal learning, reinforcement learning.

E-mail: duyu20@mails.tsinghua.edu.cn

ORCID iD: 0009-0004-3663-3130

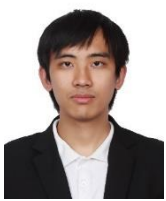

**Weihao Zhang** received the B.E. degree in software engineering from Northeastern University, Shenyang, China, in 2018. He is currently pursuing the Ph.D. degree in instrumentation science and technology with the Department of Precision Instrument, Tsinghua University, Beijing, China. His current research interests include neural network compilation, brain-inspired computing architecture, and intelligent systems.

E-mail: zwh18@mails.tsinghua.edu.cn

ORCID iD: 0000-0002-9301-8538

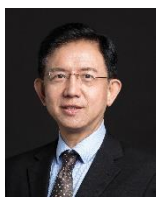

**Luping Shi** received the Ph.D. degree from the University of Cologne, Germany in 1992. During his tenure at the Data Storage Institute of the Agency for Science, Technology and Research (A*STAR) in Singapore, he served as the Director of the Artificial Cognitive Memory Lab, the Director of the Optical Materials and Systems Lab, and the Director of the Non-Volatile Memory Lab. Since joining Tsinghua University in 2013, he has been serving as the Director of the National Engineering Research Center for Optical Storage and the Director of the Tsinghua University Brain-like Computing Research Center. He is a Fellow of the International Society for Optics and Photonics (SPIE) and a Fellow of the International Institute of Cognitive Information and Cognitive Computing (I2CICC).

His current research interests include brain-inspired computing, information storage, integrated optoelectronics, intelligent systems and instruments

E-mail: lpshi@tsinghua.edu.cn

ORCID iD: 0000-0002-9829-2202

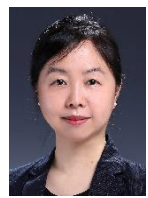

**Rong Zhao** Rong Zhao received the Ph.D. degree in electrical and computer engineering from the National University of Singapore in 1999. Currently, she is a professor at Department of Precision Instrument, the Center of Brain-inspired Computing Research and the IDG/McGovern Institute of Brain Science with Tsinghua University, China. Her research interests include brain-inspired computing and sensing, neuromorphic devices and systems, non-volatile memories and nanoelectronics. She has authored or co-authored more than 100 publications including Nature, Science, Nature Nanotechnology, Science Robotics and Nature Communications, and received 6 best papers in international conferences.

Her current research interests include low-power brain-inspired computing and chip technology, low-power, multimodal novel neural morphological devices, brain-inspired sensing devices and chip design, ultra-high-speed, low-power, high-performance non-volatile memory and chip technology.

E-mail: r_zhao@tsinghua.edu.cn （Corresponding Author）

ORCID iD: 0000-0002-2320-0326